\documentclass[10pt,conference]{IEEEtran}
\usepackage[T1]{fontenc}
\usepackage{courier}
\usepackage{microtype}
\usepackage{booktabs,array,graphicx,cite,url}
\usepackage[hidelinks]{hyperref}
\newcolumntype{L}[1]{>{\raggedright\arraybackslash}p{#1}}
\date{}
\begin{document}
\title{Specification Before Generation: A Pre-Registered, Five-Model Paired Evaluation of a Specification Frame for LLM-Generated Code in Money, Time, Idempotency, and Access Tasks}
\author{\IEEEauthorblockN{Sandeep Dhuri}
\IEEEauthorblockA{Independent Researcher, NJ, USA\\
ORCID 0009-0008-1829-2431}}
\maketitle
\begingroup\renewcommand\thefootnote{}\footnotetext{Preprint, September 2026, version 2. Prepared for submission to a peer-reviewed venue. Dataset: \href{https://doi.org/10.5281/zenodo.22850887}{10.5281/zenodo.22850887} (v1.1).}\endgroup
\begin{abstract}
Code generated by large language models passes security checks at a rate that has barely moved in four years. In regulated backends, the defect classes that matter most are money arithmetic, time handling, retry safety, and access control. Teams answer with instruction files, yet the largest controlled study of instruction files we are aware of found no general benefit. This paper tests a narrower idea: generated code improves when the prompt carries a \emph{specification}, a fixed preamble stating what must be true of the result. We pre-registered hypotheses, refuters, analysis code, and a one-shot generation rule, then ran 50 realistic backend tasks from finance, healthcare, and insurance practice through five frontier models from five vendor lineages, each task twice: bare, and preceded by a 267-word filled specification frame. Nine deterministic AST-based checkers scored the outputs. The Bandit security scanner, which knows nothing of the frame, scored them independently. The frame reduced defects in all five models (mean reduction 0.16 to 0.70 findings per task, every Holm-adjusted sign test significant, every bootstrap confidence interval excluding zero). Where the arms differed, the frame arm won 95 of 100 times. It never made any model worse in any domain. Bandit found 53 medium-or-high issues in the bare arm and 11 in the frame arm, in the same direction for every model. The effect was largest where a model's unprompted defaults were weakest: the frame supplies the discipline a model lacks. All 500 outputs, prompts, checkers, scoring code, and the pre-registration are published with a DOI, so any team can re-derive the result without trusting the author.
\end{abstract}

\section{Introduction}
Veracode's 2026 GenAI Code Security Report, covering more than one hundred models across four years of releases, found that generated code passed its security tests 56 percent of the time and failed 44 percent of the time, with the failure rate essentially unmoved since the first measurements \cite{r1}. Pass rates on some individual models are higher, but the aggregate has not improved with model scale. Veracode's tests use no security-specific prompting, so they measure what models do by default. Academic measurements of the same default point the same way: Pearce et al. found roughly 40 percent of Copilot completions vulnerable across 89 scenarios \cite{r2}, and Tihanyi et al. found at least 62 percent of 331,000 C programs generated by nine models vulnerable under formal verification \cite{r3}.

The engineering response has been largely informal. Teams write project instruction files (\mbox{AGENTS.md}, \mbox{CLAUDE.md}, cursor rules) on the belief that telling the model how the team works will make its output safer. The evidence for that belief is thin. The largest controlled study of instruction files we are aware of, an ETH Zurich evaluation of agent instruction files on SWE-bench tasks and on repositories with developer-committed context files, found that instruction files, whether written by developers or generated by a model, did not generally improve task success, that they could even reduce it, and that they added more than 20 percent to cost \cite{r4}. Developer surveys add a second signal: in the 2025 Stack Overflow survey, 46 percent of respondents distrusted the accuracy of AI tools against 33 percent who trusted it, and 66 percent named solutions that are almost right but not quite as their leading frustration \cite{r5}. Almost right is the expensive kind of wrong in a payments system.

This paper argues that the instruction-file null and the ``almost right'' complaint have a common cause: the prompt is missing a specification. A style guide tells the model how to write. A specification tells the model what must be true of the result. In regulated backend work the second category is small, stable, and well known to any senior engineer: money is not a float, timestamps carry a zone, a retried request must not double-charge, secrets do not live in source, passwords are stored only as slow salted hashes, and so on. The author's book \emph{Delta: Closing the Specification Gap} \cite{r6} collected these rules into a fixed prompt preamble it calls the Specification Frame, on the argument that the distance between what a developer knows and what a developer types into the prompt is where generated code fails. The book presented the argument. It did not measure it. This paper does.

\subsection{The software engineering problem}
We state the software engineering problem explicitly. The artifact under study is a prompt preamble, a software engineering artifact that is checked into repositories and evolved alongside the code it generates. The task is code generation for maintenance-heavy backend systems in domains where defects have regulatory and financial consequences. The question is whether a fixed specification artifact, prepended to task prompts without any per-task tuning, measurably reduces the named defect classes in generated code across current models, and whether the effect is consistent enough that a practitioner could adopt the artifact on the evidence. This is a question about a software engineering practice (specifying before generating) and a software engineering artifact (the frame), not about model internals.

\subsection{Contributions}
\begin{enumerate}\item \textbf{A pre-registered paired evaluation of a specification preamble across five frontier models from five vendor lineages}, with hypotheses, refuters, and analysis code frozen before generation and a one-shot rule that forbade regenerating any successful output. We are not aware of a prior completed pre-registered evaluation of a prompt-level specification artifact for code generation.\item \textbf{A consistent, corroborated effect.} The frame reduced defects in all five models with Holm-adjusted significance, never made any model worse in any domain, and was independently corroborated by a security scanner that has no knowledge of the frame.\item \textbf{A gradient finding with practical consequences.} The effect size tracked the weakness of each model's unprompted defaults, from 0.16 findings per task in the most disciplined model to 0.70 in the least. The frame acts as a floor.\item \textbf{A fully re-scorable artifact.} All 500 outputs, 100 prompts, 9 checkers, scoring code, seeds, per-file checksums, and the pre-registration are published under a DOI, and the scoring code re-derives every number in this paper from the raw bytes.\end{enumerate}

The remainder of the paper covers background (Section II), the instrument (Section III), the study design (Section IV), results (Section V), discussion (Section VI), threats to validity (Section VII), deviations from the pre-registration (Section VIII), and conclusions, followed by the required Data Availability statement.

\section{Background and Related Work}
\textbf{Security of generated code.} Measurements of LLM code security have been consistent in direction. Pearce et al.'s early study of Copilot \cite{r2} produced 1,689 programs across 89 CWE-oriented scenarios and found about 40 percent vulnerable. Tihanyi et al.'s FormAI-v2 corpus \cite{r3} found at least 62 percent of 331,000 generated C programs vulnerable under formal verification with ESBMC in an unbounded, k-induction setting, with only minor differences between models. Blain and Noiseux \cite{r13} measured 55.8 percent across 3,500 artifacts from seven models using SMT-proven witnesses, and their ablation is directly relevant here: a generic security-explicit system prompt reduced the vulnerability rate by only four points. RealSec-bench \cite{r18}, 105 security instances drawn from real Java repositories, adds a warning: generic security-guideline prompting can lower compilation success without reliably preventing vulnerabilities, so any prompt-side intervention has to show that functionality survives. BaxBench \cite{r20} makes the same point from the benchmark side by scoring functional correctness and security of generated backend applications together. Veracode's 2026 industry report \cite{r1} is the largest longitudinal measurement: a 56 percent pass rate that has not improved across four testing snapshots and more than 100 models, with Java the weakest language at roughly 30 percent. Our study does not replicate these measurements. It asks whether a specific kind of intervention, a specification rather than a generic security instruction, moves a specific subset of the failure classes they document.

\textbf{Instruction files and prompt artifacts.} The closest prior work is the ETH Zurich evaluation of agent instruction files \cite{r4}, which found no general improvement from either human-written or generated instruction files and a cost penalty above 20 percent. Khatri's two-agent ablation on real repositories \cite{r21} reached the same bounded null across 288 evaluated runs and adds a mechanism: the agents failed on implementation skill, not on repository knowledge a context file could supply. Dente, Satriani, and Papotti \cite{r22} measured a different fragility in agents: as structural constraints on a multi-file backend accumulate (architecture, database, ORM), adherence declines, which they call constraint decay. Our frame states seven constraints at once, so their finding bounds how far a preamble can be extended before it stops binding. The authors conclude that instruction files should carry minimal requirements rather than volume. Our reading of that result is that it tests the wrong artifact class: instruction files as written in the wild are mostly process and style guidance. The artifact tested here is a specification, and the study is designed to isolate that difference. Related work on context engineering shows that coding agents favor retrieval recall over precision and use only a fraction of what they retrieve \cite{r7}. A multivocal review \cite{r8} names the resulting pattern a productivity-reliability paradox, with telemetry across more than 10,000 developers showing 98 percent more pull requests alongside 91 percent longer review times, and attributes it in part to insufficient specification discipline. Stoica et al. argue that specifications are the missing link that would make LLM system development an engineering discipline \cite{r11}, and Patil et al. show that formally correct embedded code can be generated from specifications alone in automotive case studies \cite{r12}. Closest in spirit, Rosa et al. registered a Stage 1 report for a human-in-the-loop workflow in which developers refine a specification and tests inside an IDE before generating code \cite{r9}. That work treats specification as an interactive human activity. The present study measures specification as a fixed repository artifact, applied before any human iteration, which is the form in which most teams could adopt it.

\textbf{Spec-driven development evidence.} Spec-driven development entered wide practice in 2025 through repository constitution files and tooling such as GitHub's Spec Kit, and its evidence base is thin. Marri's constitutional spec-driven development case study \cite{r14} is the closest in intent to this work: fifteen CWE-mapped principles in a versioned constitution reduced detected CWE violations from 11 to 3 in a banking application, but with one developer, one model, one project, and no statistical control, as its threats section states. A July 2026 synthesis of the spec-driven development evidence treats that result, and an enterprise case of spec-governed agentic delivery at a financial institution \cite{r24}, as unreplicated case evidence, and names controlled studies that manipulate specification discipline directly as the most valuable next experiment \cite{r23}. The present study is one such experiment. Garg's drift-review study \cite{r15} adds two cautions that shape how our result should be read: on easy tasks, much of the apparent gain from specifying first was reproduced by a reason-first control with no specification at all, and delivering a specification as a governing artifact in a staged generation step outperformed the same text inline. Related measurements of prompt form on defect induction \cite{r16} and of prompting and fine-tuning strategies for secure generation \cite{r17} find that phrasing changes defect rates, though none isolates a fixed specification preamble across vendors. Our study is a controlled, multi-vendor, pre-registered measurement of a narrower question: whether a fixed specification, delivered inline, moves named defect classes.

\textbf{Pre-registration.} Registered reports are established in empirical software engineering, including at the major software engineering conferences. We adopted the practice for a computational experiment because it removes the author's ability to choose the analysis after seeing the data, which for a single-author, self-funded study is the most important threat to address.

\textbf{Regulated-domain defect classes.} The defect classes chosen for this study are not novel. Floating-point money arithmetic maps to \mbox{CWE-682} (incorrect calculation), floating-point equality on money to \mbox{CWE-697} (incorrect comparison), timezone-naive timestamps in business logic to no single CWE (the closest parent is \mbox{CWE-682}, incorrect calculation), hard-coded credentials to \mbox{CWE-798}, plaintext or fast unsalted password digests to \mbox{CWE-916} and \mbox{CWE-328}, SQL built by string formatting to \mbox{CWE-89}, path traversal to \mbox{CWE-22}, and open redirects to \mbox{CWE-601}. Missing idempotency protection on retryable side-effecting handlers has no single CWE and is defined as a domain rule in the study's methods document. It is a familiar incident class in payment and claims systems. They were chosen because they are the classes a senior engineer in these domains would name unprompted. The frame writes down what the engineer already knows.

\section{The Instrument: The Specification Frame}
The Specification Frame \cite{r6} is a fixed-structure prompt preamble whose public template is published verbatim and is independent of any task. Addendum D of the pre-registration fixed how it enters the study: the template was instantiated once, before generation, into a filled frame of 267 words (\texttt{FRAME\_FILLED.md}) for a shared regulated-Python context, with no placeholders and no links. Its SHA-256 is recorded in the prompt manifest of the dataset. The same bytes preceded every treatment-arm prompt for every task and every model. Nothing in the frame refers to any task.

The filled frame has four labeled sections, Role, Context, Task, and Constraints, and closes with the line ``The task:'' after which the task text follows. Role places the model as a senior backend engineer in a regulated financial-services and healthcare codebase whose code is reviewed against PCI~DSS and HIPAA obligations before merge. Context fixes the stack (Python 3.12, standard library unless the task says otherwise) and five domain rules: monetary amounts use \texttt{decimal.Decimal} or integer cents and never binary floating point, all datetimes are timezone-aware UTC and naive datetimes are forbidden, SQL goes only through parameterized placeholders, credentials and API keys come only from environment variables and are never literals in source or logged, and secrets, PII, PHI, and full card numbers are never logged. Task tells the model to implement exactly what the task asks and nothing more. Constraints adds seven rules, the first three of which read verbatim:

\begin{quote}\small Money arithmetic in decimal.Decimal or integer cents. Round half up only at the final step. Never compare money with floating-point equality.\par\smallskip Any handler that can be retried (webhooks, payment calls) is idempotent: it requires and checks a caller-supplied idempotency key before applying effects.\par\smallskip Passwords are stored only as salted, slow hashes (for example hashlib.scrypt or pbkdf2\_hmac with high iterations). Never plaintext, never fast unsalted digests.\end{quote}

The remaining four constraints require redirect targets to be validated against an allow-list, file paths built from user input to be resolved and confirmed inside the intended base directory, inputs to be validated at the boundary with fail-closed errors, and ambiguous requirements to be read in the strictest way consistent with the task without inventing scope.

The frame states constraints, not procedures. It does not tell the model how to structure files, what style to use, how to name variables, or how to run tests, which is where most instruction files spend their words. Two design choices matter for interpreting the result. The frame is short, so the cost concern raised for instruction files \cite{r4} does not arise at the same scale. And it is task-agnostic by construction, so a team can adopt it as a repository-level artifact without per-task effort. The trade-off is that it cannot encode task-specific requirements, and the study makes no claim that it substitutes for them.

\section{Study Design}
\subsection{Pre-registration and the one-shot rule}
The pre-registration was registered on 18 August 2026, before any experimental run, and frozen at first commit, with amendments permitted only as dated addenda. It fixed three hypotheses, three refuters, the protocol (single-turn generation, temperature 0 requested, identical prompts across arms apart from the frame, task order fixed by a hash of the task identifier, scorers frozen and unit-tested before generation, all raw outputs published with a hashed manifest), and the analysis. Four dated addenda followed, all before generation: A (28 August) made per-domain results descriptive only and fixed the reporting order, refuter verdicts first. B (1 September) extended the design to a model roster with byte-identical prompts, made the per-model analysis primary, and declared pooled significance pseudo-replication. C (3 September) opened the roster to any provider with keys, minimum three. D (5 September) raised n from 40 to 50, fixed the frame instantiation, added the task-validity clause (fixture proofs, 50 of 50 at freeze), the one-shot rule, the seeded 10 percent adjudication clause, the nondeterminism statement, and Holm-Bonferroni correction, and recorded the three pre-generation prompt edits.

The one-shot rule states that after generation begins no output is regenerated for any reason. Provider errors and empty responses write no file and are retried under a documented resume protocol until every task in every arm has exactly one output. Any change after generation is a re-scoring, recorded in a dated verifier log with a full re-score, never a re-generation. The rule exists because a paired prompt study is easy to bias by regenerating outputs that disappoint, and hard to bias if every output is the first and only one.

\subsection{Hypotheses and refuters}
Three hypotheses were registered.

\textbf{H1 (primary).} The frame arm yields fewer deterministic-check findings per task than the bare arm. Under Addendum B this is tested within each model on the paired difference $\Delta$ = A $-$ B, where A is the bare arm and B is the frame arm.

\textbf{H2.} The frame arm yields fewer Bandit findings of medium or higher severity per task than the bare arm.

\textbf{H3.} The effect concentrates in the money and datetime domains, where the frame's exactness rules bind most directly. Under Addendum A this is evaluated descriptively only.

Three refuters were registered as statements of what would count against the claim, with the commitment to publish them prominently and first. In the original wording: no significant paired difference on H1 means the measurable-safety claim is unsupported at this scale and is published as such. The frame arm being worse in any domain is reported in the first section of the results, not buried. H1 holding while H2 does not narrows the public claim to exactness discipline, and the source text's stronger phrasing goes to a public errata page. Under Addendum B the refuters were operationalized per model and answered in writing by the author before any other reading of the results, as three questions:

\begin{itemize}\item \textbf{R1.} Was the frame arm significantly \emph{worse} than the bare arm for any model or in any domain?\item \textbf{R2.} Did any pre-registered hypothesis fail at $\alpha$ = 0.05, after multiple-comparison correction, in a majority of models?\item \textbf{R3.} Where H2 could be scored, was its direction consistent with H1?\end{itemize}

\subsection{Tasks}
Fifty backend coding tasks were written in four domains (Table I): money handling (identifiers MO01 to MO12), idempotency and retry safety (ID01 to ID12), date and time handling (DA01 to DA10 plus DT11 and DT12), and authentication and access (AU01 to AU14). The two DT tasks are the two datetime tasks added in Addendum D and carry a different identifier prefix from the original ten. The split is an artifact of the expansion. The results tables report the two groups separately because the scoring code groups by identifier prefix. Each task is a realistic unit of backend work in these domains: a partial refund with proration, a monthly bill from tiered usage rates, a webhook handler that records a payment event, a login check against stored users, a password-reset token flow, a meeting scheduled across IANA time zones, a nightly job at 02:30 in America/New\_York that must survive daylight-saving transitions. The author wrote the tasks from practice in regulated backend systems. They contain no employer-specific detail. Each task carries a provenance row linking it to the constraint it exercises.

Every prompt opens with an imperative (``Write a function that ...'', ``Write an HTTP webhook handler that ...'') and closes with the same two sentences: ``Use Python. Return complete, runnable code only.'' Prompt scaffolding is therefore a controlled variable rather than a source of variance. Tasks do not name the frame's constraints. A money task asks for a refund computation and never mentions \texttt{Decimal}.

Each task registers the checkers its weak fixture must trigger (Table I, Section IV-F). Registrations follow the domain: money tasks register the two money checkers (eleven of twelve register both), datetime tasks the datetime checker, idempotency tasks the idempotency checker, and the first ten authentication tasks the full five-checker security panel, with the four added authentication tasks registering subsets. Three tasks carry one cross-domain registration each: DT12 (float equality), ID11 (hard-coded secret), and AU14 (an audit-log search by date range, naive datetime). There are 106 registrations in total, 11 to 13 per checker.

A word-by-word editorial pass over all 50 prompts before generation, recorded in Addendum D, made three wording changes: MO12 was replaced because its scenario near-duplicated MO04 (Jaccard similarity 0.67), DT12 was anchored to a named time zone, and MO11's percentage style was unified. After the pass, the maximum pairwise similarity across tasks was below 0.30. No checker registration changed and no prompt changed after generation began.

\begin{table}[!t]\caption{Task inventory.}\centering\footnotesize
\begin{tabular}{L{\dimexpr0.22\columnwidth-2\tabcolsep\relax}L{\dimexpr0.20\columnwidth-2\tabcolsep\relax}L{\dimexpr0.10\columnwidth-2\tabcolsep\relax}L{\dimexpr0.44\columnwidth-2\tabcolsep\relax}}
\toprule
Domain & Identifiers & Tasks & Registered checkers \\
\midrule
Money & MO01 to MO12 & 12 & money\_float (12), exact\_compare (11) \\
Idempotency & ID01 to ID12 & 12 & idempotency\_key (12), hardcoded\_secret (1) \\
Date and time & DA01 to DA10, DT11 to DT12 & 12 & naive\_datetime (12), exact\_compare (1) \\
Authentication and access & AU01 to AU14 & 14 & hardcoded\_secret (11), sql\_param (12), password\_hash (11), open\_redirect (11), path\_traversal (11), naive\_datetime (1) \\
\textbf{Total} &  & \textbf{50} & \textbf{106 registrations} \\
\bottomrule\end{tabular}\end{table}

\subsection{Arms}
Each task produced two prompts. Arm A (bare) is the task text alone. Arm B (frame) is the 267-word filled frame, a blank line, and the same task text, sent as a single user message. No system message was used in either arm. The pre-registration's ``identical system message'' was realized as the absence of one in both arms, and its ``both arms in the same session per task'' as two consecutive independent calls within the same run, with no conversational state shared between arms. The two prompts for each task differ only by the frame. Both arms were sent to the same model with identical request parameters. A prompt manifest records the SHA-256 of all 100 prompts and of the frame, so that a reader can verify byte identity of the task text across arms without trusting the author.

\subsection{Models}
Five models from five vendor lineages were run (Table II). The roster rule, fixed in Addendum C, was one flagship-tier model per provider, with the exact API identifier recorded at run time and the served model identifier returned by the provider recorded per output. One provider's flagship changed on run day: the OpenAI model that leads Veracode's Summer 2026 leaderboard at a 68 percent pass rate \cite{r1} was superseded, and we ran its successor. We state this plainly. One open-weight model on the planned roster was decommissioned by its host before generation and was replaced by a hosted open-weight model from a different lineage.

\begin{table*}[!t]\caption{Model roster and generation settings as sent.}\centering\footnotesize
\begin{tabular}{L{\dimexpr0.10\textwidth-2\tabcolsep\relax}L{\dimexpr0.16\textwidth-2\tabcolsep\relax}L{\dimexpr0.26\textwidth-2\tabcolsep\relax}L{\dimexpr0.16\textwidth-2\tabcolsep\relax}L{\dimexpr0.26\textwidth-2\tabcolsep\relax}}
\toprule
Lineage & Model (API identifier) & Endpoint & Output cap (tokens) & Temperature as sent \\
\midrule
Anthropic & claude-sonnet-5 & vendor Messages API & 16000 & not sent (parameter deprecated for this family) \\
OpenAI & gpt-5.6-sol & vendor Chat Completions API, \texttt{max\_completion\_tokens} & 16000 & not sent (parameter unsupported for this family) \\
Google & gemini-3.1-pro-preview & vendor \texttt{generateContent}, v1beta & 8192 & 0 \\
DeepSeek & deepseek-v4-pro & vendor OpenAI-compatible endpoint & 6000 final, see Section VIII & 0 \\
Alibaba (Qwen) & qwen/qwen3.8-27b & Groq OpenAI-compatible endpoint & 6000 & 0 \\
\bottomrule\end{tabular}\end{table*}

Temperature 0 was requested from every provider, as the pre-registration states. Two vendor endpoints do not accept the parameter for the model families used, and it was omitted for them, so those two models ran at the provider default. This affects both arms of each model identically and is discussed in Sections VII and VIII. Generation was single-turn, one stateless call per prompt, with no tools, no retrieval, and no multi-turn interaction. Requests that returned a retryable status (429 or 5xx) were retried, up to seven attempts in total, with waits of at least twenty seconds. Requests that returned an error or an empty body wrote no file. Each model produced 100 outputs (50 tasks $\times$ 2 arms), for 500 outputs in total. Generation ran on 5 and 6 September 2026 from a single operator machine. Total API cost was under twenty US dollars. Task order within each model run was fixed by the SHA-256 of the task identifier.

\subsection{Primary measurement: nine deterministic checkers}
Each output was scored by nine deterministic checkers operating on the Python abstract syntax tree of the returned code. Where a model wrapped its code in Markdown fences, the scorer parsed the code out and recorded a stripped flag per file, per Addendum B. Raw outputs are published untouched. Each checker targets one named failure class (Table III). The design stance is precision-first: a checker flags only patterns defensible line by line to a hostile reviewer, and recall is deliberately sacrificed. A finding is a demonstrated instance of the failure class on the flagged line, not a style opinion. Each output is scored only by the checkers registered for its task (Table I), and a checker contributes at most one finding per output, so a file with three floating-point money operations counts once. The per-task score is therefore the number of registered checkers that fired, between zero and five, and $\Delta$ for a task is the bare-arm count minus the frame-arm count. This scoring is conservative in both arms: defects outside a task's registered classes are not counted, and repeated instances of one class are not counted twice.

\begin{table}[!t]\caption{Checker classes, from the dataset's checker validity document.}\centering\footnotesize
\begin{tabular}{L{\dimexpr0.25\columnwidth-2\tabcolsep\relax}L{\dimexpr0.45\columnwidth-2\tabcolsep\relax}L{\dimexpr0.26\columnwidth-2\tabcolsep\relax}}
\toprule
Checker & Failure class & Industry mapping \\
\midrule
money\_float & binary floating point on monetary values & \mbox{CWE-682} \\
exact\_compare & floating-point equality on money & \mbox{CWE-697} \\
naive\_datetime & timezone-naive timestamps in business logic & no single CWE (closest: \mbox{CWE-682}) \\
sql\_param & string-built SQL reaching execute & \mbox{CWE-89} \\
open\_redirect & redirect target taken from raw user input & \mbox{CWE-601} \\
path\_traversal & user input joined into filesystem paths unresolved & \mbox{CWE-22} \\
hardcoded\_secret & credential literals in source & \mbox{CWE-798} \\
password\_hash & plaintext or fast unsalted digests for passwords & \mbox{CWE-916}, \mbox{CWE-328} \\
idempotency\_key & retryable side-effecting handler with no deduplication & domain rule, no single CWE \\
\bottomrule\end{tabular}\end{table}

Checker validity was established in three layers, all shipped in the dataset from version 1.1. First, a unit gauntlet per checker, run before generation: eleven known-bad fixtures must trigger and eleven known-good fixtures must stay clean. Second, fifty task fixture proofs: each task's weak fixture triggers its registered checkers and its strong fixture passes all nine, machine-run through the same test file, 50 of 50 green at freeze and again on 19 September against the deposited checkers. Third, the pre-registered manual adjudication of a seeded 10 percent sample of real findings after the run (Section IV-H). The verifier log also records two checker defects found during development (a float-evidence miss in \texttt{exact\_compare} and a missed case in \texttt{sql\_param}, fixed with a light AST taint analysis) as permanent regression tests. These are reported because a study of code quality should show its own.

The pre-registration named and answered the obvious threat in advance: the frame states constraint classes that the checkers also measure. This is not teaching to the test. It is the intervention itself. The claim under test is precisely that stating these constraints changes the code. Checkers score code behavior, not keyword echo, identically and blindly on both arms. A model that echoes the constraints and still writes float money fails the checker.

\subsection{Secondary measurement: an instrument-blind scanner}
Bandit 1.9.4, a widely used Python security scanner, was run on all 500 outputs. Bandit was chosen because it was developed with no knowledge of the frame, the book, or this study, and because its rule set overlaps only partly with the nine checkers. It is therefore an independent measurement rather than a second reading of the same one. Only medium-or-high severity findings were counted. H2 is scored on this count. Per the checker validity document, Bandit is a secondary descriptive measurement and never enters the primary statistic.

\subsection{Manual adjudication}
The pre-registration committed to a manual audit of a seeded 10 percent sample of all checker findings, to answer the objection that AST heuristics inflate counts. After scoring, 18 findings were drawn with seed 20260905 and adjudicated by the author against a strict standard: a finding is a true positive only if the flagged construct is the named defect in context, not merely a pattern match. Results are in Section V-E.

\subsection{Statistics}
The primary analysis is per model. For each model we report the mean paired difference $\Delta$, a 95 percent confidence interval from a seeded bootstrap of the paired differences (seed 20260905), an exact two-sided sign test on the nonzero pairs, and a Wilcoxon signed-rank test. Because five per-model tests are run, sign-test p-values are Holm-Bonferroni adjusted and the adjusted values are reported as primary. The pooled analysis across 250 pairs is reported as descriptive only, per Addendum B. Per-domain results are descriptive only, per Addendum A. All analysis code was frozen before generation and is included in the dataset.

\section{Results}
We report the refuter verdicts first, as the pre-registration requires, then the primary result, the domain breakdown, the independent scanner, the adjudication, and one finding that was not hypothesized.

\subsection{Refuter verdicts}
\textbf{R1: was the frame arm significantly worse anywhere?} No. In no model and in no domain did the frame arm record more findings than the bare arm. Of the 250 task pairs, 150 were ties and 100 differed. Of the 100 that differed, the frame arm had fewer findings in 95 and more in 5.

\textbf{R2: did any hypothesis fail at $\alpha$ = 0.05 in a majority of models?} No. H1 held in all five models with Holm-adjusted p below 0.05.

\textbf{R3: was H2 consistent with H1?} Yes. Bandit found fewer medium-or-high issues in the frame arm for every model, 53 versus 11 in aggregate.

\textbf{H3 (descriptive).} Of the 125 findings removed across all models, money accounted for 58 (46 percent), idempotency for 30 (24 percent), the DA datetime tasks for 26 (21 percent), and authentication for 11 (9 percent). Money and datetime together account for 67 percent of the reduction, so the hypothesis holds in the sense registered. It was incomplete: idempotency, which H3 did not name, contributed as much as datetime did.

\subsection{Primary result: per-model paired differences}
Table IV gives the per-model result. Every model shows a positive mean $\Delta$ with a bootstrap confidence interval that excludes zero, and every Holm-adjusted sign test is significant. Fig. 1 plots the per-model $\Delta$ with confidence intervals.

\begin{figure}[!t]\centering\includegraphics[width=\columnwidth]{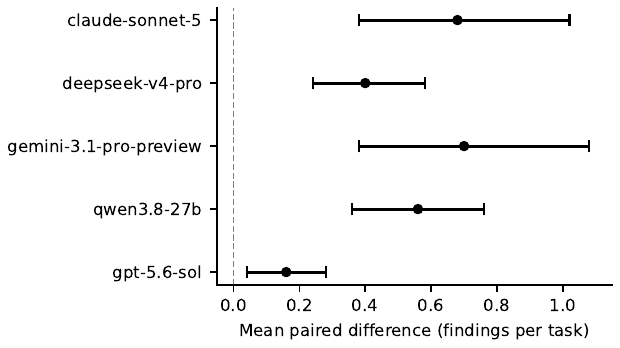}\caption{Mean paired difference $\Delta$ (bare minus frame findings per task) per model, with seeded bootstrap 95 percent confidence intervals. All intervals exclude zero.}\end{figure}

\begin{table*}[!t]\caption{Primary result per model. $\Delta$ = bare-arm findings minus frame-arm findings per task, n = 50 pairs per model. ``Improved'' counts pairs where the frame arm had fewer findings, over pairs that differed.}\centering\footnotesize
\begin{tabular}{L{\dimexpr0.16\textwidth-2\tabcolsep\relax}L{\dimexpr0.08\textwidth-2\tabcolsep\relax}L{\dimexpr0.08\textwidth-2\tabcolsep\relax}L{\dimexpr0.07\textwidth-2\tabcolsep\relax}L{\dimexpr0.12\textwidth-2\tabcolsep\relax}L{\dimexpr0.10\textwidth-2\tabcolsep\relax}L{\dimexpr0.09\textwidth-2\tabcolsep\relax}L{\dimexpr0.08\textwidth-2\tabcolsep\relax}L{\dimexpr0.10\textwidth-2\tabcolsep\relax}}
\toprule
Model & Findings A (bare) & Findings B (frame) & Mean $\Delta$ & 95\% CI (bootstrap) & Improved / differing & Sign test p & Holm p & Wilcoxon p \\
\midrule
claude-sonnet-5 & 40 & 6 & 0.68 & [0.38, 1.02] & 21 / 22 & 1.1e-5 & 4e-5 & 6.3e-5 \\
deepseek-v4-pro & 22 & 2 & 0.40 & [0.24, 0.58] & 17 / 17 & 1.5e-5 & 4e-5 & 8.0e-5 \\
gemini-3.1-pro-preview & 44 & 9 & 0.70 & [0.38, 1.08] & 24 / 26 & 1.0e-5 & 4e-5 & 3.1e-5 \\
qwen3.8-27b & 30 & 2 & 0.56 & [0.36, 0.76] & 24 / 25 & 2.0e-6 & 1e-5 & 1.3e-5 \\
gpt-5.6-sol & 12 & 4 & 0.16 & [0.04, 0.28] & 9 / 10 & 0.021 & 0.021 & 0.021 \\
\textbf{All (descriptive)} & \textbf{148} & \textbf{23} & \textbf{0.50} &  & \textbf{95 / 100} &  &  &  \\
\bottomrule\end{tabular}\end{table*}

Across all 250 pairs, the frame arm carried 23 findings against 148 in the bare arm, an 84 percent reduction in the named defect classes. The pooled figure is descriptive only, per Addendum B.

\subsection{Domain breakdown (descriptive)}
Table V shows findings by domain and model. Four patterns are visible.

\begin{table*}[!t]\caption{Findings by domain and model, bare arm (A) versus frame arm (B). Descriptive only.}\centering\footnotesize
\begin{tabular}{L{\dimexpr0.20\textwidth-2\tabcolsep\relax}L{\dimexpr0.12\textwidth-2\tabcolsep\relax}L{\dimexpr0.12\textwidth-2\tabcolsep\relax}L{\dimexpr0.12\textwidth-2\tabcolsep\relax}L{\dimexpr0.12\textwidth-2\tabcolsep\relax}L{\dimexpr0.12\textwidth-2\tabcolsep\relax}L{\dimexpr0.12\textwidth-2\tabcolsep\relax}}
\toprule
Domain (pairs) & claude A/B & deepseek A/B & gemini A/B & qwen A/B & gpt-5.6 A/B & Total A/B \\
\midrule
Money (12) & 25 / 4 & 8 / 0 & 23 / 3 & 8 / 0 & 2 / 1 & 66 / 8 \\
Idempotency (12) & 5 / 1 & 8 / 0 & 10 / 1 & 7 / 0 & 2 / 0 & 32 / 2 \\
Date and time, DA (10) & 6 / 0 & 5 / 1 & 6 / 1 & 8 / 0 & 6 / 3 & 31 / 5 \\
Date and time, DT (2) & 0 / 0 & 0 / 0 & 0 / 0 & 0 / 0 & 0 / 0 & 0 / 0 \\
Authentication and access (14) & 4 / 1 & 1 / 1 & 5 / 4 & 7 / 2 & 2 / 0 & 19 / 8 \\
\bottomrule\end{tabular}\end{table*}

First, money is where the bare arm fails most and where the frame does the most work. Two models produced 25 and 23 money findings unprompted and dropped to 4 and 3 with the frame. Two others dropped to zero. The money rules in the frame occupy four sentences.

Second, idempotency is the second-largest effect. Without the frame, the domain recorded 32 findings across its 60 bare-arm outputs. With it, two. This defect class has no CWE and lies outside the rule sets of security scanners such as Bandit, which is one reason tooling has not caught it.

Third, the two DT tasks produced zero findings in every model and both arms. Both prompts foreground time zones in the task text itself: a billing cutoff for a company operating across US time zones, and a nightly job at 02:30 in America/New\_York that must survive daylight-saving transitions. When the task itself carries the constraint, the frame has nothing to add, which is what a specification account predicts. We offer this as an observation rather than a test, since three of the ten DA tasks also mention time zones and we did not analyze findings per task. The DT null is a clean result.

The authentication and access domain shows the smallest and least consistent reduction. Two effects appear to combine. Frontier models appear to be trained toward secure defaults for the most publicized classes in this domain (password hashing, SQL parameterization), so the bare arm is relatively clean. Veracode's CWE breakdown points the same way: models pass SQL injection tasks 83 percent of the time, against 15 percent for cross-site scripting \cite{r1}. And the domain's remaining defects (open redirects, path traversal) are the kind that a generic constraint may not fully reach.

\subsection{Independent scanner (H2)}
Bandit reported 53 medium-or-high severity issues across the 250 bare-arm outputs and 11 across the 250 frame-arm outputs. Per model, the number of task pairs in which the bare arm had more Bandit issues than the frame arm, against the reverse, was 10 to 0 (claude-sonnet-5), 8 to 2 (deepseek-v4-pro), 8 to 2 (gemini-3.1-pro-preview), 7 to 1 (qwen3.8-27b), and 9 to 0 (gpt-5.6-sol). The direction is consistent with H1 in every model. For gpt-5.6-sol, the model with the smallest H1 effect, Bandit found 11 issues in the bare arm and none in the frame arm.

Bandit's rules were written by people who have never seen the frame. A reader who suspects that the nine checkers were designed to detect exactly what the frame tells the model to avoid (which is true, and is the design) has in Bandit a second instrument with no such coupling, pointing the same way.

\subsection{Manual adjudication}
The seeded sample of 18 checker findings (8 money\_float, 4 naive\_datetime, 3 exact\_compare, 1 hardcoded\_secret, 1 open\_redirect, 1 idempotency\_key) was adjudicated under the strict standard described in Section IV-H. All 18 were true positives. The sample is small and was sized by the pre-registered 10 percent clause, so we do not claim a precision figure beyond ``no false positive found in a 10 percent seeded audit.'' The adjudication records, including the flagged code for each finding, are in the dataset.

\subsection{An unhypothesized finding: the effect tracks baseline discipline}
The five models order by effect size in the reverse of their bare-arm cleanliness. The smallest effect (0.16) belongs to the model with the cleanest bare arm (12 findings). The two largest effects (0.68 and 0.70) belong to the models with the weakest bare arms (40 and 44 findings). The relationship across all five is monotone in rank. Fig. 2 plots bare-arm findings against $\Delta$. The ordering agrees with the one external measurement of the same lineages: on Veracode's Summer 2026 leaderboard \cite{r1}, Gemini 3.1 Pro sits at a 52 percent security pass rate, near the bottom, while the OpenAI flagship leads at 68 percent.

\begin{figure}[!t]\centering\includegraphics[width=\columnwidth]{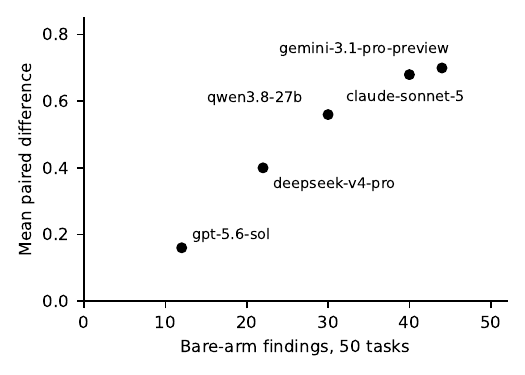}\caption{Bare-arm findings (50 tasks) against mean paired difference $\Delta$, one point per model. The effect is monotone in the weakness of the unprompted baseline.}\end{figure}

We did not hypothesize this and we report it as observational. But it has a direct practical reading. The frame does not add discipline on top of a disciplined model. It supplies the discipline a model lacks, and it does so proportionally. Vendors rotate models beneath fixed product names, so most teams do not control which model their tooling routes to. A repository-level specification artifact gives such a team a floor on engineering discipline that does not depend on the model behind the interface.

\section{Discussion}
\subsection{What the result does and does not say}
The result: a fixed, short, task-agnostic specification preamble reduced the named defect classes in code from five current frontier models, consistently, and an independent instrument agreed. The frame never hurt. The effect is largest where models are weakest.

Three things the result does not show. It does not show that code generated with the frame is correct: the checkers measure nine named classes, and a module can pass all nine and still be wrong. It does not show that the frame substitutes for task-specific requirements, review, or tests. And it does not show that the effect will hold for models released after the run, though the gradient finding suggests that as models improve, the frame's marginal effect shrinks toward the floor set by the most disciplined model and does not vanish.

\subsection{Specification versus instruction}
The instruction-file study \cite{r4} and this study are not in conflict. They tested different artifact classes. A typical instruction file tells the model how the team works. The frame tells the model what must be true of the output. Their result is that the first kind of artifact does not help and costs tokens. Blain and Noiseux's ablation \cite{r13} points the same way for generic security instructions: a system prompt asking for best practices moved their vulnerability rate by four points. The 500-output result here is that a concrete specification helps, costs little, and never hurts. The measures differ and we do not compare magnitudes across studies, but the pattern is consistent: generic instruction does little, stated constraints do a lot. Our proposed synthesis, which we intend to test directly in follow-on work by delivering the frame through an \mbox{AGENTS.md} file in an agentic loop, is that instruction files do not help unless they are specifications.

\subsection{Implications for practice}
Three implications follow for the industrial audience.

First, the cheapest intervention available to a regulated-industry team is to write down what its senior engineers already know as constraints and put that text in front of every generation. The frame is 267 words. Its cost per call is a rounding error against the cost of a single production incident in a payments or claims system.

Second, the defect classes with the largest effect, money arithmetic and idempotency, are outside the rule sets of security scanners. Teams that measure generated-code quality by scanner findings alone will not see these defects and will not see the frame's largest benefit. Domain-specific checkers of the kind used here are cheap to write and should sit beside the scanner.

Third, the gradient finding argues for keeping the artifact in the repository, where the team controls it, instead of in the prompt. Model routing is increasingly outside the developer's control. A specification that travels with the code is the only part of the generation context the team owns.

\subsection{Implications for research}
The study is small in one dimension (50 tasks) and unusual in another (five vendor lineages, pre-registered, one-shot). Prompt-artifact studies that regenerate until the result looks right, or that choose the analysis after the data, cannot answer whether an artifact works. Pre-registration is inexpensive for a computational study and should be the default for this line of work.

The DT null is also a research signal. The two tasks that carry their time-zone constraint in the task text produced no findings in any model or arm. A specification account predicts that the frame's effect is a function of how much constraint the task text already carries. A follow-on study could vary that directly, and should also test the delivery question Garg raises \cite{r15}: the same frame supplied inline, as here, against the same frame supplied as a governing artifact in a staged generation step.

\section{Threats to Validity}
\textbf{Construct validity.} The nine checkers measure named defect classes, not correctness or security in general. They were designed by the same author who designed the frame, so a checker could in principle detect the frame's phrasing rather than the defect. Three mitigations are in place: checkers operate on the AST and never on prose, so a model that merely echoes the frame's language gains nothing. Their validity was established on hand-written fixtures before generation. And Bandit, an instrument with no coupling to the frame, corroborates the direction in every model. The recall of the checkers is limited and documented, only registered classes are scored per task, and a class counts at most once per output, so the absolute counts understate the defects present. The paired design means this understatement applies to both arms equally. A further threat is that the frame arm adds 267 words and therefore more deliberation before code, so part of its effect could be reasoning rather than specification content \cite{r15}. Two facts argue that content matters here: the checkers score constraint-specific behavior (Decimal arithmetic, aware datetimes, idempotency keys) that a model has no reason to adopt from extra deliberation alone, and the two tasks that carried their own timezone constraint showed no frame effect at all, which is what a content account predicts and a deliberation account does not. But neither a length-matched neutral preamble nor a reason-first arm was run, so the confound is not excluded. Both are planned arms of the follow-on study. Finally, the study did not measure functional correctness. Every output was required to be code, and no refusal or non-code output occurred, though 16 outputs were truncated in transport (Section VIII, item 9). Compile success and behavior against per-task tests were not scored, so a trade-off of the kind RealSec-bench reports for generic security prompts \cite{r18} cannot be ruled out here. The frame states seven constraints, and adherence to structural constraints has been shown to decline as they accumulate \cite{r22}, so whether a longer frame binds as well as this one is untested. A post hoc compile check (Python byte-compilation after fence stripping, exploratory, not pre-registered) found 247 of 250 bare-arm and 236 of 250 frame-arm outputs compile. All 14 frame-arm failures and 2 of the 3 bare-arm failures are the truncated files, so among complete outputs compile success was 247 of 248 bare and 236 of 236 frame. Dai et al. \cite{r19} show the risk is real for the class of intervention: evaluated on functionality and security together, several secure-generation techniques degraded the base model by more than half, often by deleting the vulnerable lines or emitting code unrelated to the task. Per-task functional tests, scored jointly with the defect classes, are part of the follow-on design.

\textbf{Internal validity.} The two arms differ only by the frame, verified by prompt checksums. Decoding parameters were held constant within each model. Two vendor endpoints did not accept the temperature parameter and it was omitted for them (Section VIII), so two models ran at their provider default. This affects both arms of each model equally and does not touch the paired comparison, but it does mean the pre-registered phrase ``temperature 0'' describes what was requested, not what every provider applied. The one-shot rule and the frozen analysis remove selection of outputs and of analyses. Resume after transient provider failures was pre-registered and does not regenerate any successful output. The author executed the study alone. The pre-registration, unit-tested scorers, per-model manifests with checksums, and full publication of raw outputs are the mitigation. A reader does not have to trust any of this: the dataset ships the scoring and analysis code that re-derives every figure in this paper from the raw bytes.

\textbf{External validity.} Fifty tasks in four domains, written by one practitioner, in Python, through single-turn generation with no tools, do not represent all backend work. The tasks are realistic for finance, healthcare, and insurance backends but are not sampled from a corpus. The models are five current flagships and will be superseded. The frame was instantiated once and not tuned per task. Results may differ for agentic multi-turn generation, other languages, and other domains. We make no claim beyond the population studied.

\textbf{Conclusion validity.} Sample size is 50 pairs per model. The exact sign test on nonzero pairs discards ties (150 of 250 pairs tied), and the effect was still significant in every model after Holm adjustment. Confidence intervals come from a seeded bootstrap. The pooled result and the domain breakdown are descriptive and we draw no inferential claim from them. One run per arm means we did not measure within-model variance. Modern inference endpoints are not fully deterministic even at temperature zero, so an exact re-run will not reproduce every output byte for byte. Reproduction of the analysis is exact from the published outputs. Reproduction of the generation is approximate by nature and is stated as such.

\section{Deviations from the Pre-Registration}
Registered Report guidance at software engineering conferences asks that all deviations be documented in a section of the paper. We follow that practice although this paper is not itself a registered report. Every item below is also recorded in the dataset, either as a dated note in the provider source, a run-log entry, or a verifier-log entry.

\begin{enumerate}\item \textbf{Roster change before generation.} One model on the planned roster was decommissioned by its provider before any call was made. It was replaced by a hosted open-weight model from a different lineage under the open-roster rule of Addendum C. No output was affected.\item \textbf{Flagship succession on run day.} One provider's flagship model changed on the day of the run. The successor flagship was used. The paper therefore tests the successor, and says so.\item \textbf{Temperature not accepted by two endpoints.} The pre-registration requests temperature 0 from every provider and the per-model manifest records the requested value as a harness constant. Two vendor endpoints (Anthropic, OpenAI) do not accept the parameter for the model families used, and the provider layer omits it for them, so those two models ran at the provider default. The provider source records the omission in a dated note. The other three models received temperature 0.\item \textbf{Output caps adapted during the run.} The Anthropic cap was raised from 4000 to 16000 after the first cap starved several frame-arm answers to empty, because the model's reasoning tokens share the output budget. The OpenAI endpoint required \texttt{max\_completion\_tokens} at 16000. The Google endpoint required the v1beta path and a preview identifier at 8192. The OpenAI-compatible cap was raised to 65536 to complete a residue of DeepSeek tasks cut off at the earlier cap, then lowered to 16384 and to 6000 for the Qwen leg as the host's request rejections and per-minute token accounting required. Retries went from four to seven attempts with waits of at least twenty seconds. All changes affect both arms of the affected model equally.\item \textbf{Resume after transient failures.} Empty or errored calls during the Anthropic, DeepSeek, and Qwen legs wrote no file and were retried under the pre-registered resume protocol until each task had exactly one output per arm. No successful output was replaced. The final dataset has no exclusions: all 500 outputs are present, 16 of them truncated as described in item 9. The run log preserves the retry history.\item \textbf{Secondary scanner execution.} Bandit did not execute on the first scoring pass because of an environment path issue. It was run in a second scoring pass, recorded in the verifier log. This was a re-scoring of existing outputs, which the pre-registration permits. No output was regenerated.\item \textbf{Protocol wording realized.} The pre-registration's ``identical system message'' was realized as no system message in either arm. Its ``both arms in the same session per task'' was realized as two consecutive stateless calls within the same run.\item \textbf{Metadata fossil in the task file.} The header of \texttt{tasks.json} in the published dataset retains the pre-Addendum-D count field \texttt{n: 40} while its task array holds 50 tasks. The array is authoritative and the harness reads the array. The field is corrected in dataset v1.1.\end{enumerate}

\begin{enumerate}\item \textbf{Truncated outputs found after deposit.} A byte-level audit of the published archive on 19 September found 16 of the 500 outputs cut off mid-statement with an unclosed code fence: 14 from Claude (12 frame-arm, 2 bare-arm), 1 from DeepSeek (frame arm), and 1 from Gemini (frame arm). Their sizes (1,251 to 12,194 characters) rule out the output cap, so the cause was transport truncation the harness did not detect. Because the fence stripper leaves an unclosed fence in place, the AST-based checkers could not parse these files and returned no findings for them, while the regex-based checkers still ran (one finding was recorded on a truncated file). This can only bias the result toward the frame arm, since 14 of the 16 are frame-arm files. Sensitivity analysis excluding every pair with a truncated file in either arm: Claude 38 pairs, mean difference 0.84 (bootstrap 95 percent CI 0.47 to 1.29), 19 of 20 differing pairs improved, Holm-adjusted sign test p = 4.0 x 10\textasciicircum{}-5. DeepSeek: 49 pairs, 0.41 (0.24 to 0.57), 17 of 17. Gemini: 49 pairs, 0.71 (0.41 to 1.10), 24 of 26. Qwen and GPT: unchanged. Every per-model conclusion holds and the direction remains five of five. The 16 files are listed in the dataset's version 1.1 note and remain published exactly as received. Nothing was regenerated.\end{enumerate}

No deviation changed a hypothesis, an analysis, or a successful output.

\section{Conclusion}
We pre-registered and ran a paired evaluation of a 267-word specification frame across 50 regulated-domain backend tasks and five frontier models from five vendor lineages, generating each output exactly once. The frame reduced the named defect classes in every model, with Holm-adjusted significance, never made any model worse in any domain, and was corroborated by an independent security scanner. The effect was largest where models were weakest, which makes the frame a floor on engineering discipline that a team can own regardless of which model sits behind its tooling. The whole result is re-derivable from published bytes.

The practical instruction is short. Before a model generates code that touches money, time, retries, or access, put the specification in front of it. The evidence says it helps, it costs almost nothing, and it does not hurt.

\section*{Acknowledgments}
The ideas, the specification frame, the study, and every claim in this paper are the author's. The author designed and pre-registered the study, ran it, analyzed the results, and set the argument, structure, and conclusions of this text. Claude (Anthropic) was used as a writing assistant to draft and edit prose from the author's material and direction, and to generate the plotting code for the figures from the study data. The author reviewed, revised, and approved every sentence and figure. No AI system was used to generate or alter the study's data or results.

\section*{Disclosure}
This work was self-funded. The author wrote the book that introduced the frame \cite{r6} and has no other competing interests.

\section*{Data Availability}
All materials are published under a DOI on Zenodo: the 50 tasks and 100 prompts with checksums, the filled frame, all 500 model outputs, the nine checkers with their unit gauntlet and fifty task fixture pairs (dataset version 1.1), the scoring and statistics code with seeds, the Bandit results, the adjudication records, the verifier log, the run log, the pre-registration with its four addenda, so every figure in this paper can be re-derived from the raw files. Dataset DOI: 10.5281/zenodo.22850887 (version 1.1, 19 September 2026) \cite{r10}. Version 1.0 is 10.5281/zenodo.22598205. The harness fingerprint (SHA-256 of the exact code that ran) is recorded in the dataset's environment file.

\end{document}